# Stochastic biophysical modeling of irradiated cells

Krzysztof Wojciech Fornalski *

* PGE EJ 1 Sp. z o.o., Technology and Operations Office, ul. Mysia 2, 00-496 Warszawa, Poland
e-mail: krzysztof.fornalski@gmail.com, krzysztof.fornalski@gkpge.pl

### ABSTRACT

The paper presents a computational stochastic model of virtual cells irradiation, based on Quasi-Markov Chain Monte Carlo method and using biophysical input. The model is based on a stochastic tree of probabilities for each cell of the entire colony. Biophysics of the cells is described by probabilities and probability distributions provided as the input. The adaptation of nucleation and catastrophe theories, well known in physics, yields sigmoidal relationships for carcinogenic risk as a function of the irradiation. Adaptive response and bystander effect, incorporated into the model, improves its application. The results show that behavior of virtual cells can be successfully modeled, e.g. cancer transformation, creation of mutations, radioadaptation or radiotherapy. The used methodology makes the model universal and practical for simulations of general processes. Potential biophysical curves and relationships are also widely discussed in the paper. However, the presented theoretical model does not describe the real cells and tissues. Also the exposure geometry (e.g., uniform or non-uniform exposure), type of radiation (e.g., X-rays, gamma rays, neutrons, heavy ions, etc.) as well as microdosimetry are not presently addressed. The model is focused mainly on creation of general and maximal wide mathematical description of irradiated hypothetical cells treated as complex physical systems.

## 1. INTRODUCTION

An influence of ionizing radiation on single cells or cell colonies can be described by many deterministic and stochastic models. All such models can be easily implemented as computational algorithms, as shown for example in (UNSCEAR 1986; Moolgavkar and Luebeck 1990; Feinendegen *et al.* 2000, 2010; Brenner *et al.* 2001; Calabrese and Baldwin 2003; Feinendegen 2005; Scott *et al.* 2007, 2013; Leonard 2008; Shuryak *et al.* 2009; Bogen 2011; Leonard *et al.* 2011a, 2011b; Scott 2011, 2013; Tavares and Tavares 2013; Wodarz *et al.* 2014). In 2011 the fully stochastic approach using the Markov Chain Monte Carlo method was implemented to the virtual cells colony (Fornalski *et al.* 2011a), and some additional modifications were published a few months later (Fornalski *et al.* 2011b). In the following paper a significant development of the model proposed by (Fornalski *et al.* 2011a, 2011b) is presented.

In the former model (Fornalski *et al.* 2011a, 2011b) a group of virtual cells were irradiated and many simplified biophysical mechanisms were implemented using the Markov Chain Monte

Carlo stochastic method. Many popular effects, like adaptive response and bystander effect, were successfully accounted for. The results showed that presence of multiple relationships linearly depending on the dose results in the overall reaction which is generally non-linear. The final curves showing a fraction of mutated and cancerous cells can be supported by experimental data (Fornalski *et al.* 2011b). Because the model (Fornalski *et al.* 2011a, 2011b) consists of a tree of probabilities, all known biophysical effects can be accounted for, which makes the model very flexible in use. A prospective user of the model can add new branches to the probability tree and/or select a completely different set of input probabilities. It allows for easy expansion of the model to more sophisticated forms of the relationship between probabilities and doses.

The main differences between the new version of the model (described in this paper) and previous ones is the implementation of time-dependent relationships and changeable values of doses during the life of a virtual cell colony. In addition, all probability distribution relationships describing the biophysics of the virtual cell are usually continuous and differentiable, which is necessary to describe natural processes.

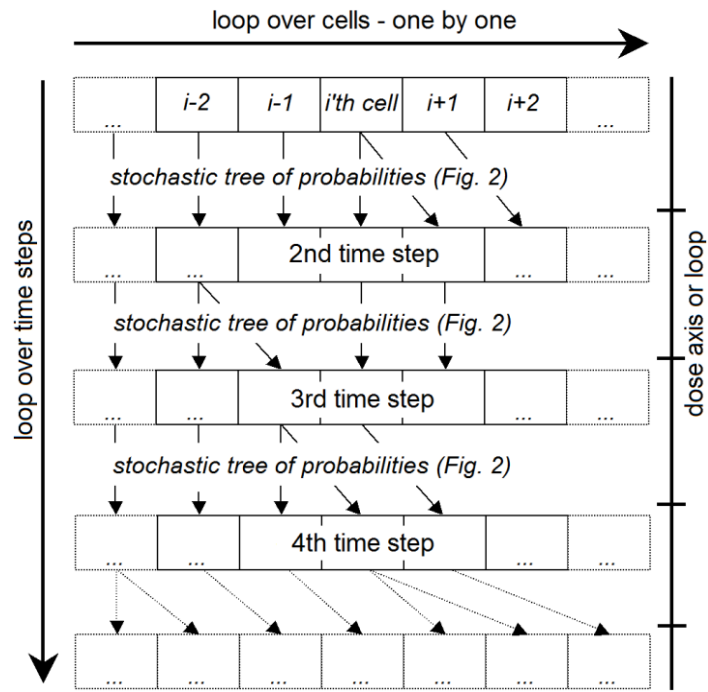

Figure 1. The general scheme of the proposed algorithm. In each time step i-th cell is let through the tree of probabilities (Fig. 2). Moreover in each time step the dose can be different.

Table I. The description of variables used in the text

| Variable | Description |
|---|---|
| K | The time step. In general situation the variable $K$ corresponds to the time |
| D | The dose in single time-step. This concept corresponds to the dose-rate. Formally, the cumulative dose equals $K \cdot D$ |
| Q | Number of mutations in the cell |
| ξ | The general variable equals $K$, $D$ or $Q$ |
| N | The number of cells at the beginning of the algorithm |
| S | The status of the cell: healthy ($S_H$), mutated ($S_M$) or cancerous ($S_C$) |
| i | The index of actually analyzing cell |
| τ, δ, a | Scaling parameters |
| n | Index parameter |
| β₁, β₂ | Parameters used in bystander effect description |
| α₁, α₂, α₃ | Parameters used in adaptive response effect description |

## 2. ALGORITHM

The model can be easily simulated using computer algorithm in any programming language. The algorithm is composed of a few numerical loops nestled one in another (Fornalski *et al.* 2011a, 2011b), e.g. the loop over radiation dose, the loop over the time steps and the loop over the cells (Fig. 1). In this context all cells are computationally arranged as a one dimensional multivariable chain (Fig. 1). In a single time step ($K$) one has a single dose pulse ($D$) and a numerical loop over all cells in the presented virtual organism/colony. Taking cell-by-cell into account, it is easy to use a tree of probabilities (Fig. 2) to get a new state of each cell (healthy/mutated/cancer/death one).

The most important and novel aspect of the model is changeable value of dose[1] ($D$) in each single time step ($K$), see Tab. I. When the value of $D$ is constant over time steps, it can be analogically treated as a dose ratio[2]. Thus, the total cumulative dose is equal to $D \cdot K$.

The tree of probabilities, displayed in Fig. 2, contains several input parameters, like probability distributions of many biophysical effects. The tree can be easily modified and new branches can be added according to better knowledge or just to focus on one detailed mechanism in cell.

The algorithm starts with $N$ cells, which can have different status (healthy/mutated/cancer), but usually at the beginning they are healthy ones. The age of each initial cells is taken randomly[3]. The first loop (over all $N$ cells) starts and for each i-th cell the three of probabilities is used (Fig. 1). At the beginning, the Boolean status of cell ($S$) is checked (Fig. 2). Cell can be healthy ($S_H$), mutated ($S_M$) or cancerous ($S_C$). Status $S_M$ means that the cell have minimum one damage in DNA, including chromosomal aberrations (Renan 1993), while $S_C$ means that cell is tumorigenically

---

[1] Dose $D$ mean the absorbed dose measured in the not-defined general unit UAD (Unit of Absorbed Dose), which is proportional to Gy (Gray)

[2] The conversion of the "dose-rate" into "number of dose-pulses" with varying time intervals in between them, can be very useful in the context of numerical simulation. This important generalization conforms to the understanding of dose-rate in microdosimetry terms

[3] Taken from the discrete uniform distribution with minimum and maximum of [1,50]

transformed[4] (Elmore *et al.* 2005). After checking status of the i-th cell (Fig. 2), it can be irradiated (hit by the ionizing particle) with the probability $P_{hit}$ and new six states are considered: healthy-irradiated ($S_H \cdot P_{hit}$), healthy-nonirradiated ($S_H \cdot (1-P_{hit})$), mutated-irradiated ($S_M \cdot P_{hit}$), mutated-nonirradiated ($S_M \cdot (1-P_{hit})$), cancer-irradiated ($S_C \cdot P_{hit}$) and cancer-nonirradiated ($S_C \cdot (1-P_{hit})$).

Then, next branches arise from these six states:

a)  when the healthy cell is irradiated, it can:

- naturally die (probability $P_D$),

- be killed by a single radiation hit ($P_{RD}$),

- spontaneously mutate, irrespectively of the hit ($P_M$),

- naturally multiply ($P_S$),

- generate bystander signal to the nearby cell ($P_B$),

- become mutated by radiation ($P_{RM}$),

- stay intact ($1-P_D-P_{RD}-P_S-P_M-P_B-P_{RM}$);

b)  when the healthy cell is not irradiated, it can:

- naturally die (probability $P_D$),

- spontaneously mutate ($P_M$),

- naturally multiply ($P_S$),

- stay intact ($1-P_D-P_S-P_M$);

c)  when the mutated cell is irradiated, it can:

- naturally die (probability $P_{MD}$),

- be killed by one precise hit of radiation ($P_{RD}$),

- spontaneously mutate one more time ($P_M$),

- naturally multiply ($P_{MS}$),

- generate the adaptive response effect ($P_A$),

- generate a bystander effect ($P_B$),

- spontaneously repair one of the existing mutations ($P_R$),

- be mutated by radiation ($P_{RM}$),

- stay intact ($1-P_{MD}-P_{RD}-P_{MS}-P_M-P_A-P_B-P_R-P_{RM}$);

d)  when the mutated cell is not irradiated, it can:

---

[4] Which means that the cell become a cancerous one

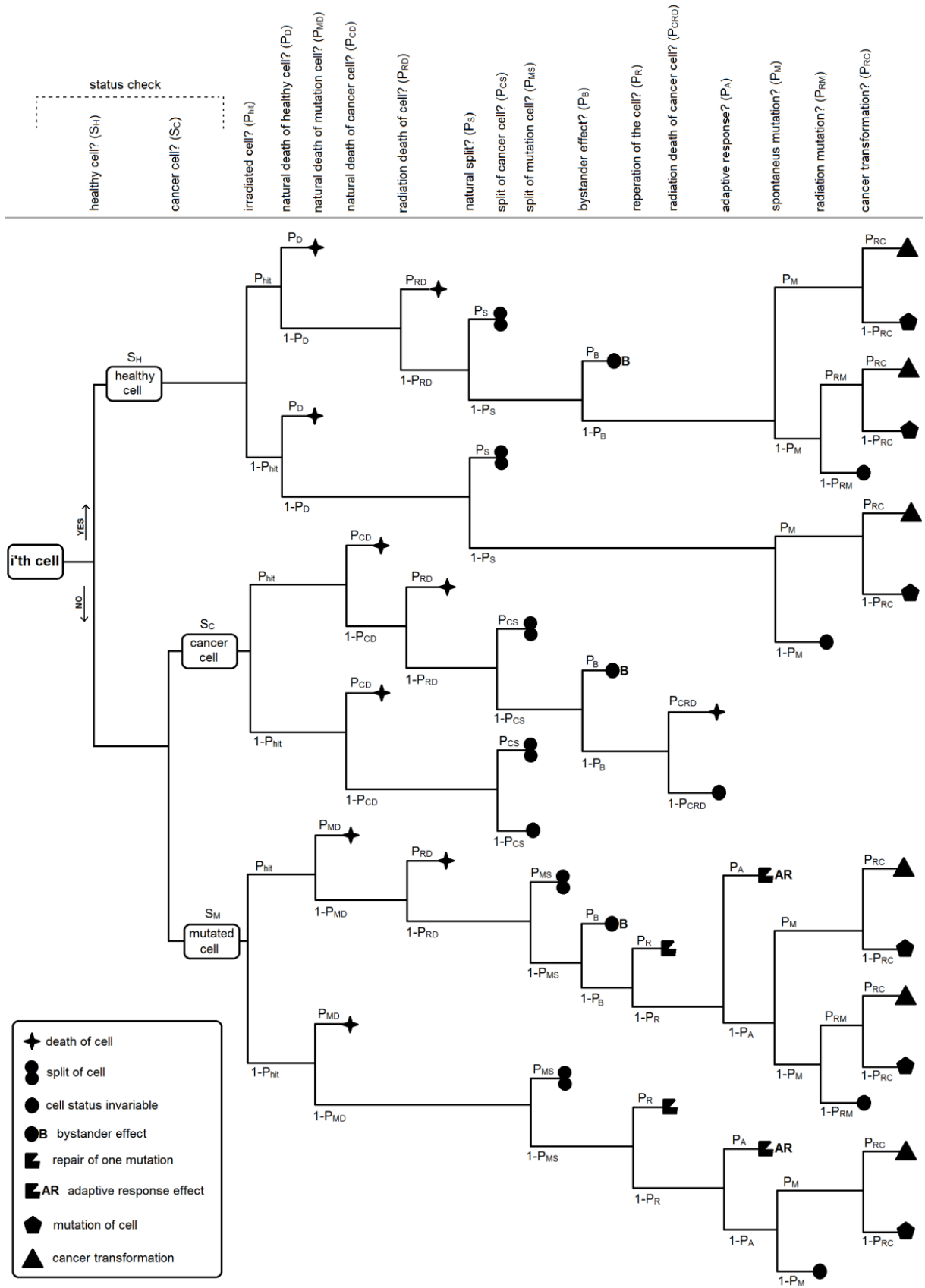

Figure 2. The tree of probabilities implemented to each cells in each time steps (Fig. 1). After traversing (letting through) the tree, the cell can change the status (healthy/mutated/cancer/death one) or not.

- naturally die (probability $P_{MD}$),

- spontaneously mutate one more time ($P_M$),

- naturally multiply ($P_{MS}$),

- generate an adaptive response signal ($P_A$),

- naturally repair one existing mutation ($P_R$),

- stay intact ($1-P_{MD}-P_{MS}-P_M-P_A-P_R$);

e) when the cancer cell is irradiated, it can:

- naturally die (probability $P_{CD}$),

- be killed by one precise hit of radiation ($P_{RD}$),

- naturally multiply ($P_{CS}$),

- die because of the radiation-induced additional damage due to cancer cell's radiosensitivity ($P_{CRD}$),

- generate a bystander effect ($P_B$),

- stay intact ($1-P_{CD}-P_{RD}-P_{CS}-P_{CRD}-P_B$);

f) when the cancer cell is not irradiated, it can:

- naturally die (probability $P_{CD}$),

- naturally multiply ($P_{CS}$),

- stay intact ($1-P_{CD}-P_{CS}$).

In the situation, where the branch with the new mutation appears ($P_M$ or $P_{RM}$), the cell is let through the next ramification and the probability of cancer transformation ($P_{RC}$) is used. The whole algorithm displayed in Fig. 2 was presented graphically also in (Fornalski *et al.* 2011a).

### 3. PHYSICAL BASIS AND INPUT DATA

The computer algorithm was written using C++ computational language with an object-oriented program overlay and library, named ROOT[5]. It uses a table-simulated virtual colony of cells while all general biophysical processes are implemented as probability distributions (probability density functions) used in stochastic tree of probabilities (Fig. 2). The next iteration over all time steps brings results of a Monte Carlo simulation (Fig. 1). Because in many cases the recent state of i-th cell is independent of previous ones, one can say that those process are quasi-Markov ones (only time-dependent distributions, $P_D$, $P_{MD}$, $P_M$, $P_R$, $P_A$, break the Markov rules) (Podgórska *et al.* 2002).

All probabilities presented in previous section form input data. The potential user of the model can use their own values of probabilities or probability distributions, according to the best

---

knowledge of cell's biophysics. This is what makes the model rather flexible. All proposed model's input probabilities and probability distributions are presented below.

### 3.1. Fixed probabilities

Some of the probabilities presented in the previous section can be approximated by simple constant values, e.g.:

- probability of natural death of cancer cell ($P_{CD}$);

- probability of the healthy cell dividing ($P_S$); the newly formed cell (daughter cell) is also healthy, as well as the mother cell;

- probability of dividing of mutated cell ($P_{MS}$); the newly formed cell (daughter cell) is also mutated and have the same number of mutations ($Q$) as the mother cell;

- probability of dividing of cancer cell ($P_{CS}$); the newly formed cell (daughter cell) is also cancerous, as well as the mother cell.

The exact values of probabilities $P_{CD}$, $P_S$, $P_{MS}$ and $P_{CS}$ can be taken from experimental results if available, otherwise they may be simulated or taken arbitrarily. Remaining probabilities presented in previous section need to be described in the form of probability distributions.

### 3.2. Quasi-linear probability distributions saturating at large doses

The most common and useful form of quasi-linear probability distribution dependent on dose ($D$) is (Bryszewska and Leyko 1997; Simmons and Watt 1999; Leonard 2008; Scott *et al.* 2013):

$$P(\xi) = 1 - e^{-const \cdot \xi} \tag{1}$$

where $\xi = D$. For low doses the equation (1) is quasi-linear and tends asymptotically to 1 at large doses. The equation (1) can also be used for:

- probability of hitting a cell with radiation ($P_{hit}$) – the equation (1) can be treated as an equivalent of the notion of cross section used, e.g. in particle physics, because radiation is a stream of ionizing particles; the equation (1) is also used as a "hit probability" in (Leonard 2008; Scott *et al.* 2013);
- probability of new mutation induced by radiation ($P_{RM}$) – for the healthy cell it would be the first mutation (status change from healthy to mutated); for mutated cell the number of existing mutations $Q$ will increase by one; the $P_{RM}$ probability distribution is connected with the so called "(quasi)linear model" proposed in (Bogen 2011; Scott 2013);
- probability of sudden death of cell induced of by precise hit of radiation ($P_{RD}$), independent of the cell's status;
- probability of death of irradiated cancer cell because of its specific radiosensitivity ($P_{CRD}$).

### *3.3. Sigmoidal probability distributions – Avrami-type equations*

Each cell in the algorithm can have few states: healthy, mutated (with $Q$ mutations), cancerous as well as dead. The mutated cell means that the DNA chain contains some damage, such as chromosomal aberrations, displacements or deletions. The existence of mutation is one of the causes of tumorigenic transformation into a cancer cell (Elmore *et al.* 2005), which is connected with the $P_{RC}$ probability distribution. The process of creation of a cancer cell can be described in many ways. One of them stems from a theory of nucleation and growth (Avrami 1940, 1941; Cohen *et al.* 2011; Sinha and Mandal 2011), catastrophe theory (Zeeman 1977; Cobb and Watson 1980; Aerts *et al.* 2003) or theory of self-organised criticality (Stark 2012). The common idea underlying these theories is a cumulative impact of some environmental stressors (here: radiation) on complex adaptive systems that may result in a rapid non-linear response when the stress exceeds some critical value. As M. Stark wrote in the context of sandpile model (Stark 2012): "*Too much stress* [here: the radiation – author annotation] – *traumatic stress – will be too overwhelming for the system to manage, triggering instead devastating breakdown* [here: the cancer transformation – author annotation]. *Too little stress will provide too little impetus for transformation and growth, serving instead simply to reinforce the system's status quo.*". In the context of the theory of nucleation and growth (Avrami 1940, 1941), the cancer creation can be treated as a rapid non-linear growth appearing around the germ. Within the scope of this theory, the sigmoid function with a critical index $n$, describing the dimensionality of the growth, is dependent on number of mutations ($Q$), and the probability distribution for cancer transformation, $P_{RC}$, may be described in the form:

$$P(\xi) = 1 - e^{-a\xi^n} \qquad (2)$$

where $\xi = Q$. The sigmoidal eq. (2) is a so called Avrami equation (Avrami 1940, 1941) or Johnson-Mehl-Avrami-Kolmogorov (JMAK) equation (Cohen *et al.* 2011; Sinha and Mandal 2011). In general, the variable $\xi$ can equal $D$, $K$ or $Q$ in the case of the situation[6].

The probability that the mutated cell is transformed into the cancerous one ($P_{RC}$) can be described by equation (2). The probability of tumorigenic transformation is dependent on the number of mutations in the cell, $Q$, which can be generally described by the Knudson hypothesis (Nordling 1953; Knudson 1971). It has been shown in experimental kinetic analyses of human cell cultures that four to seven rate-limiting stochastic events – thought to be distinct somatic mutations ($Q$) and the triggered cellular signaling pathways thereof – are required for the formation of a tumorigenic cell (Renan 1993; Hahn *et al.* 1999; Hahn and Weinberg 2002). Based on this assumption, one can model the probability of transformation of a mutated cell into a cancerous cell, $P_{RC}$, using e.g. the equation (2).

It should be pointed out, however, that the process described above intentionally omits: (1) evidence consistent with the requirement for only two mutations (Armitage and Doll 1957; Moolgavkar 1988), (2) observations that adult cancer-cell genomes typically contain 1000 to 10,000 somatic mutations that arise in a very non-homogenous manner over time (Stratton 2011; Stephens *et al.* 2011), as well as (3) evidence consistent with a limited role if

---

[6] The general variable $\xi$ was used due to the fact, that many equations are more general than to use for $D$ or $Q$ only

any for mutations *per se* in tumorigenesis (Bogen 2013). However, it is always possible to adapt a double-hit or multiple-hit model (Armitage and Doll 1957; Ashley 1969; Armitage 1985; Moolgavkar 1988; Moolgavkar and Luebeck 1990) instead of proposed simplified eq. (2).

More general form of equation (2) is a scaled sigmoid function with additional scaling factor τ:

$$P(\xi) = (1 - \tau) \cdot (1 - e^{-a\xi^n}) + \tau \tag{3}$$

This form of sigmoid can well describe functions which starts from non-zero point, e.g. *P(0) = τ*. The variable *ξ=K* is typically a number of time steps (the age of cell), so the form (3) can be dependent on time.

The scaled sigmoid function (3) for *ξ=K* can approximate following probability distributions, which are time-dependent:

- probability that a healthy cell dies because of natural reasons ($P_D$);
- probability that a mutated cell dies because of natural reasons ($P_{MD}$);
- probability of a spontaneous mutation in a cell (irrespective of irradiation) – first in a healthy cell, and the following in a mutated one ($P_M$).

The probability of natural repair of one mutation ($P_R$) in the mutated cell is also time dependent, but the possibility of repair does not increases but decreases with time (aging of the cell). Such process can be described by an inverted sigmoid function, e.g.:

$$P(\xi) = \delta\, e^{-a\xi^n} \tag{4}$$

where *ξ=K*. The probability $P_R$ described by the equation (4) intends to describe repair of one mutation[7] in the mutated cell[8]. In the case of just one mutation in the cell, status of repaired cell changes into the healthy one.

The sigmoidal Avrami's function proposed in presented model can be described in simple form (eq. (2)) or in scaled forms (eq. (3) and (4)) for *ξ={K,Q,D}*. However, the sigmoid function is wider mathematical concept, so the equations (2)-(4) are not only ones. There are also some other possibilities, for instance the so called linear-quadratic concept (Bogen 2011)

$$P(\xi) = 1 - c_1\, e^{-c_2\,\xi - c_3\,\xi^2} \tag{5}$$

or logistic one (Dabrowski and Thompson 1998; Bogen 2011):

$$P(\xi) = \frac{e^{c_1 + c_2\xi}}{c_3 + e^{c_1 + c_2\xi}} = \frac{c_4}{c_5 + c_6 e^{-c_7\xi}} \tag{6}$$

All of relationships (eq. (2)-(6)) are sigmoidal functions, but (2) and (3) are preferred in the model according to the nucleation theory, especially for $P_{RC}$. Each sigmoid function has a property called the inflection point (or cliff effect), at which the steepest variation is observed. The inflection point can be calculated from

$$\frac{d^2 P(\xi)}{d\xi^2} = 0 \qquad (7)$$

and for the most general form of eq. (3) is given by

$$\xi_{inflection\_point} = \sqrt[n]{\frac{(n-1)(1-\tau)}{a\,n\,(1-\tau^2)}} \qquad (8)$$

which can be very useful for experimental evaluation of the sigmoid function shape. In the case of simpler eq. (2) and planar growth of tumor ($n=2$) this value is $(2a)^{-1/2}$.

---

[7] The repair of one mutation is a numerical simplification of a real, much complicated, process
[8] However, in general situation with many repair acts, the $P_R$ relationship is usually characterized as having a half-time implicating more repair events with increasing time; see the point entitled *Repair of damages* in *Discussion* section

The special form of sigmoid function is a quasi-linear one described by eq. (1). In this case $n = 1$, which simply means non-critical linear relationship. This quasi-linear function, as well as original sigmoid one, can be found in nature. For example the probability distribution describing the hit probability $P_{hit}$, eq. (1) (Leonard 2008; Scott $et\ al.$ 2013), can also be found in the context of cross section ($\sigma$) in particle and nuclear physics.

### 3.4. Bystander effect

The bystander effect (Leonard 2008; Brenner $et\ al.$ 2001) can appear with the probability $P_B$ dependent on dose, $D$. Suggestion of its functional shape, similar to equation (1), but with additional scaling parameter, was given by (Prise $et\ al.$ 2003):

$$P(\xi) = \beta_1\left(1 - e^{-\beta_2 \xi}\right) \qquad (9)$$

where $\xi = D$. The bystander effect can cause mutations in cells neighboring an irradiated cell. The proposed algorithm randomly selects one nearby cell from the investigated i-th cell. When the neighboring cell is healthy its status will change from "healthy" to "mutated" (assuming one mutation in the cell). The number of mutations in this nearby cell increases by 1 if the cell had earlier mutations. The investigated i-th cell preserves its status (and the number of mutations). Notably, in this approach it was assumed that the bystander effect causes only adverse outcomes and its possible beneficial effects (Prise $et\ al.$ 2003; Leonard 2008; Mothersill and Seymour 2006) have been ignored. Within the present context of the algorithm a positive result would enhance the branch of the adaptive response (see the next subsection).

### 3.5. Adaptive response effect

Potential beneficial effects due to induced excessive repair of radiation damage has been described within the context of adaptive response, as described by (Feinendegen $et\ al.$ 2000; Feinendegen 2005; Leonard 2008). The hormetic effect, strictly connected with adaptive response, is typically described by the parabolic equation (Fornalski $et\ al.$ 2012)

$$P(\xi) = a - b\,\xi + c\xi^2 \qquad (10)$$

for $\xi = D$. The *NOAEL* point (*No Observed Adverse Effect Level*, (Calabrese and Baldwin 1993; Fornalski *et al.* 2012)) is observed at

$$NOAEL = \frac{b + \sqrt{b^2 - 4ac}}{2a} \tag{11}$$

and maximal protection factor (*PROFAC*, (Scott 2005; Sanders and Scott 2008)) for $\xi = \frac{1}{2} b/c$ equals

$$PROFAC = \frac{b^2}{4c} \tag{12}$$

However, equations (10)-(12) are not universal ones, because $\lim_{\xi \to \infty} P(\xi) = \infty$ for eq. (10). For that reason one can postulate the convenient saturated hormetic distribution given e.g. as a scaled relationship[9] (Bogen 2011):

$$P(\xi) = (1 - \tau)(1 - c_1 \, e^{\, c_2 \, \xi - c_3 \, \xi^2}) + \tau \tag{13}$$

The *NOAEL* and *PROFAC* values can also be written in more general forms as

$$NOAEL = \frac{c_2 + \sqrt{c_2^2 - 4c_3 \ln \frac{1}{c_1(1-\tau)}}}{2c_3} \tag{14}$$

and (for $\xi = \frac{1}{2} c_2/c_3$)

$$PROFAC = 1 + c_1(\tau - 1) \, e^{\frac{c_2^2}{4c_3}} \tag{15}$$

respectively. Equations (14) and (15) are useful for experimental evaluation of general hormetic curve (eq. (13)).

The specific general shape of hormetic function (eq. (13)) may differ from case to case, nevertheless it offers convenient qualitative description of the benefit effect. In order to model this effect one has to describe the probability of the adaptive response effect (Leonard 2008), $P_A$. It appears only in mutated cells and reduces the number of mutations in the cell by 1 (model's numerical assumption). If the cell has only one mutation, its status will change from "mutated" to "healthy". Probability of the adaptive response $P_A$ should be given by probability distribution function with the maximum value at low doses as well as with the strongest effect appearing after some period of time (Feinendegen 2005). The probability distribution of adaptive response should thus be dependent both on the dose ($D$) and time ($K$).

The dose and time ingredients of adaptive response can be approximated by following distributions, respectively:

$$P(D) = \alpha_1 D^\nu e^{-\alpha_2 D} \tag{16}$$

$$P(K) = \alpha_4 K^\mu e^{-\alpha_3 K} \tag{17}$$

---

[9] At small doses both hormetic equations (10) and (13) can be used alternatively using approximated conversion variables as $a = \tau + (1 - \tau)(1 - c_1)$, $b = (1 - \tau)c_1 c_2$ and $c = (1 - \tau)c_1 c_3$.

The proposed distributions of eq. (16) and (17) are rather general and they are dose or time dependent, separately. To make the distributions more precise, the leading constants have to be chosen so to properly normalize probability distributions according to experimental data. Here the emphasis is put on the general shapes of the dose and time dependent reactions, so the exact values of multipliers are not relevant at the level of pure calculations. The choice of exponents $\nu$ and $\mu$ is also somewhat arbitrary, but they should be not less than 1 to create the adaptive-like shape of both curves[10].

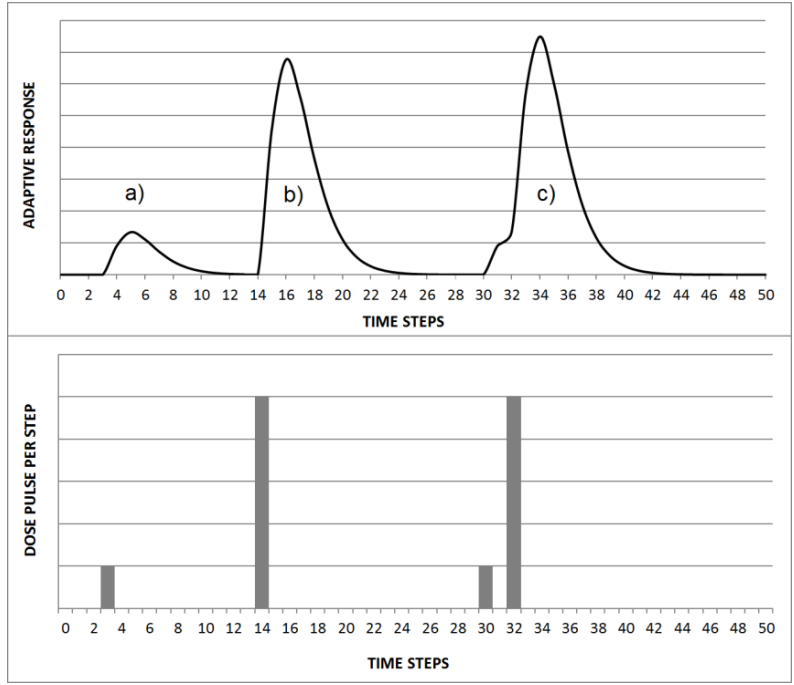

Figure 3. Simulated examples of adaptive response effect proposed in eq. (21). Example a) shows single small pulse, b) single but stronger pulse and c) shows additive effect of two pulses in the short period of time.

To merge eq. (16) and (17) into one dose and time dependent equation, the maximal value of probability distribution (17) have to be determined by the actual value of $D$ for probability (16). Another words: after cell's irradiation by the dose $D$, one can find the probability connected with this value of $D$ (16). This probability is a maximal value of the time-dependent probability distribution (17). In that way the time-dependent adaptive response distribution can be strictly connected with the dose formerly received.

To make eq. (17) dependent on eq. (16), one have to find the maximum of distribution (17), $P(K_0)$, where $K_0$ can be calculated from $\frac{dP(K)}{dK} = 0$, and put it into (16):

$$P(K_0) = P\left(\frac{\mu}{\alpha_3}\right) = \alpha_4 \left(\frac{\mu}{\alpha_3}\right)^{\mu} e^{-\mu} \equiv \alpha_1 D^{\nu} e^{-\alpha_2 D} \qquad (18)$$

---

[10] In the other way the desired hunchbacked shape of the adaptive response curve would disappear

Compiling equation (18) with (17) using constant $\alpha_4$, one can find the time and dose dependent distribution:

$$P(D,K) = \alpha_1 \left(\frac{\alpha_3}{\mu}\right)^\mu D^\nu K^\mu e^{\mu - \alpha_2 D - \alpha_3 K} \qquad (19)$$

The time- and dose-dependent distribution described by eq. (19) is almost complete. One can assume also $\nu = \mu = 2$ for simplicity (Fornalski *et al.* 2011b). However, the eq. (19) describes the adaptive response from one single irradiation only. In real situation, one can find the different values of $D$ during next time steps $K$. For this reason it is necessary to find the adaptive response signal extended over time coming from $n$ single doses $D_i$:

$$P_A \equiv \sum_{i=1}^n \sum_{j=0}^{K_{max}} P(D_i, K_j) \qquad (20)$$

and finally

$$P_A = \sum_{i=1}^n \sum_{j=0}^{K_{max}} \frac{\alpha_1 \alpha_3^2}{4} D_i^2 K_j^2 e^{2 - \alpha_2 D_i - \alpha_3 K_j} \qquad (21)$$

where $K_j$ is an actual time step from the moment of receiving dose $D_i$. In other words: each dose $D_i$ in fact generates one signal (19) extended over time, which is additive to rest *n-1* signals. In practical computational algorithm, it is necessary to create a multidimensional table for each cell. The table contains all information such as status, all $n$ doses $D_i$ received by cell, the time steps $K_j$ passed from each $D_i$ and one summarized signal $P_A$ generated from each $D_i$.

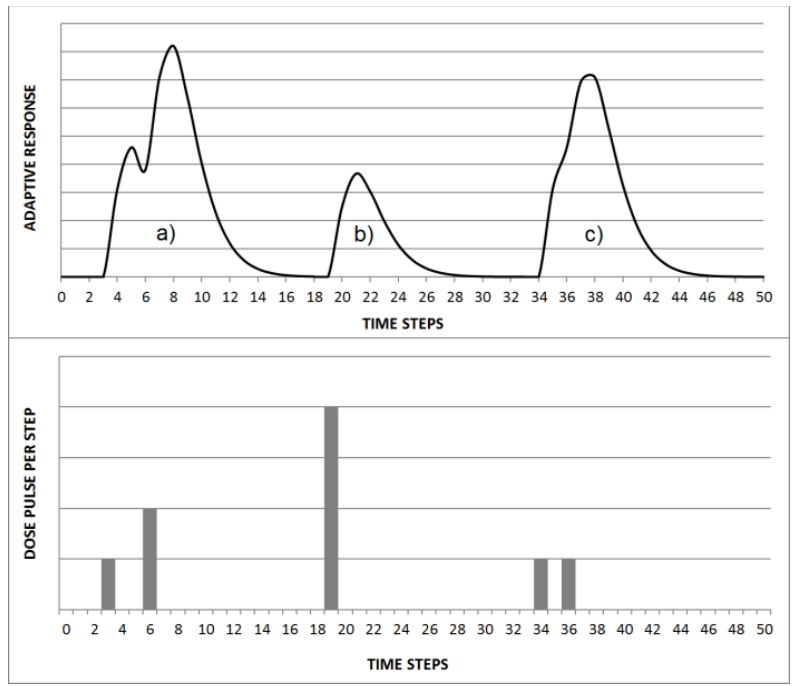

Figure 4. Simulated examples of adaptive response effect proposed in eq. (21). Example a) shows two small pulses (the higher is the same like in Fig. 3b), b) single but too strong pulse (adaptive response is relatively smaller) and c) shows additive effect of two small pulses in the short period of time.

Proposed dose and time dependent adaptive response equation (21) would be a fairly universal one. The adaptive response signal can be simulated according to single dose pulses, additional dose pulses, constant dose-rate etc. Some examples of equation (21) application are presented in Fig. 3. The example a) shows a single small dose pulse (lower panel) and adaptive response signal (upper panel). Example b) shows similar but stronger pulse separated from the previous one by the long time period. Figure 3c presents two pulses from examples a) and b) but the time between them is rather short. One can see the additive effect and relatively stronger adaptive response. It is necessary to add, that the pulse cannot be too strong, because such pulse would rather decrease than increase adaptive response effect, like in Fig. 4b. The higher dose pulse from Fig. 4a is the same as in Fig. 3b. An interesting example is presented in Fig. 4c, where two relatively small pulses received briefly can amplify the adaptive response effect. Similar situation is presented in Fig. 5, where dose pulses are the same over time (constant dose-rate). In this situation the adaptive response saturates at a constant value.

The specific shape of adaptive response function (eq. (21)) can be treated as the so called benefit function (Scott 2013).

All probabilities and equations presented above and chosen to be used in the model, are summarized in Tab. II. Moreover, Tab. II contains propositions of exemplary values of all input parameters. Those parameters can also be taken from recent experimental results or assumed arbitrarily (see Discussion for details).

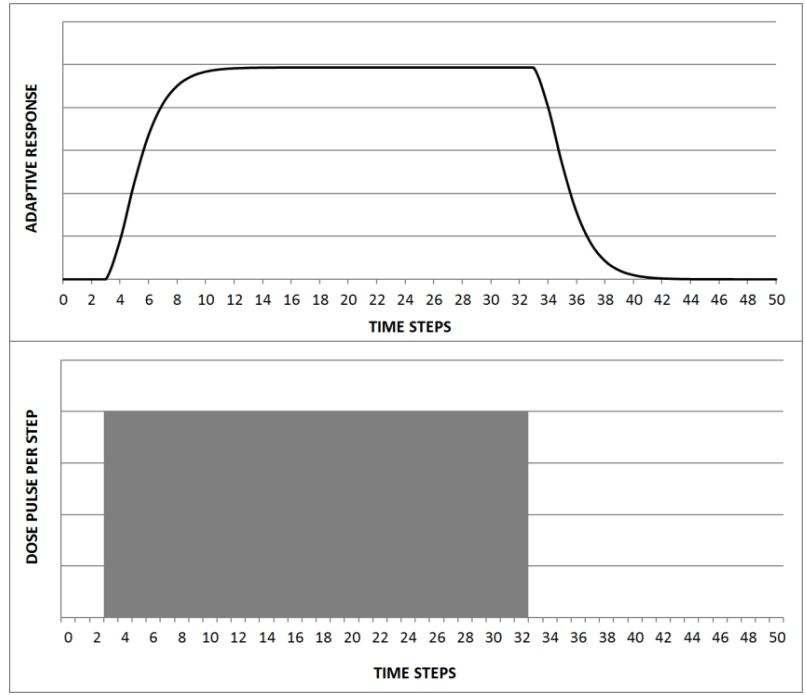

Figure 5. Simulated example of adaptive response effect proposed in eq. (21). Example shows a bundle of identical dose pulses (constant dose-rate) and saturated adaptive response.

Table II. The summary of all input parameters, including probabilities and probability
distributions with exemplary values of all variables. See also Tab. I for description of variables

| Probability | Equation | Exemplary values | Description |
|---|---|---|---|
| $P_{hit}$ | $1 - e^{-const \cdot D}$ | $const = 0.04$ | Probability that a cell is hit (irradiated) by a particle of radiation |
| $P_D$ | $(1-\tau) \cdot (1 - e^{-aK^n}) + \tau$ | $\tau = 0.002$ $a = 3 \cdot 10^{-7}$ $n = 3$ | Probability that a healthy cell dies because of natural reasons |
| $P_{MD}$ | $(1-\tau) \cdot (1 - e^{-aK^n}) + \tau$ | $\tau = 0.003$ $a = 7 \cdot 10^{-7}$ $n = 3$ | Probability that a mutated cell dies because of natural reasons |
| $P_{CD}$ | $const$ | $const = 0.001$ | Probability that a cancer cell dies because of natural reasons |
| $P_S$ | $const$ | $const = 0.02$ | Probability that a healthy cell multiplies |
| $P_{MS}$ | $const$ | $const = 0.01$ | Probability that a mutated cell multiplies |
| $P_{CS}$ | $const$ | $const = 0.03$ | Probability that a cancer cell multiplies |
| $P_M$ | $(1-\tau) \cdot (1 - e^{-aK^n}) + \tau$ | $\tau = 10^{-4}$ $a = 10^{-8}$ $n = 3$ | Probability of a spontaneous mutation in a cell |
| $P_R$ | $\delta\, e^{-aK^n}$ | $\delta = 10^{-4}$ $a = 10^{-5}$ $n = 4$ | Probability of natural repair of one mutation in the mutated cell |
| $P_B$ | $\beta_1 (1 - e^{-\beta_2 D})$ | $\beta_1 = 0.001$ $\beta_2 = 300$ | Probability of a bystander effect |
| $P_A$ | $\sum_{i=1}^{n} \sum_{j=0}^{K_{max}} \frac{\alpha_1 \alpha_3^2}{4} D_i^2 K_j^2 e^{2 - \alpha_2 D_i - \alpha_3 K_j}$ | $\alpha_1 = 1$ $\alpha_2 = 100$ $\alpha_3 = 1$ | Probability of an adaptive response |
| $P_{RM}$ | $1 - e^{-const \cdot D}$ | $const = 0.5$ | Probability that a mutation develops in the irradiated cell |
| $P_{RC}$ | $1 - e^{-aQ^n}$ | $a = 10^{-4}$ $n = 5$ | Probability that the mutated cell is transformed into the cancerous one |
| $P_{RD}$ | $1 - e^{-const \cdot D}$ | $const = 6 \cdot 10^{-5}$ | Probability that a cell dies from a precise hit by radiation |
| $P_{CRD}$ | $1 - e^{-const \cdot D}$ | $const = 0.003$ | Probability that a cancer cell dies because of its specific radiosensitivity |

## 4. STOCHASTIC RESULTS

The input parameters with proposed probability distributions from Tab. II were used for the algorithm presented in Fig. 2. The simulations demonstrated the stochastic behavior of a group of cells in many different circumstances, like time-evolution of single colony when dose is varied or dose-evolution of many colonies, exposed in constant dose-rate. In the following some exemplary results are described in detail.

### 4.1. Constant dose-rate

The easiest case to test the present model is the simulation of a behavior of $N$ healthy cells (assumed to be 1000 at the beginning) irradiated with a constant dose-rate[11] during 500 time steps. The time evolution of single group of cells is obtained using the input parameters listed in Table II. Fig. 6 shows examples of healthy cells growth when the dose-rate is constant for each example. The value of constant dose-rate for each example changes from 0 UAD (Unit of Absorbed Dose) to 50 UAD per time step. In the case of 0 UAD/step (no radiation) the regular growth of the healthy cells is observed. However, when the radiation level exceeds 0.2 UAD/step, the number of healthy cells start decreasing and more and more mutated cells appear (not shown in Fig. 6).

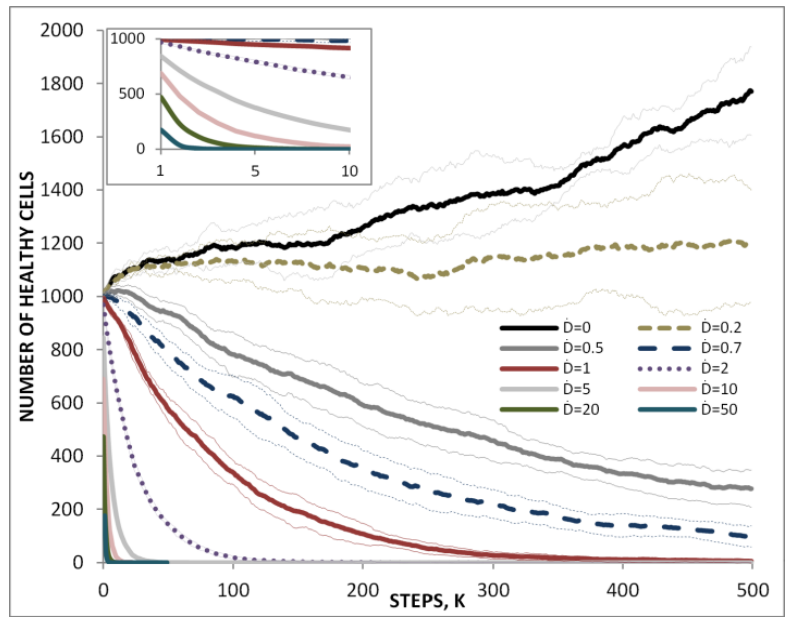

Figure 6. The number of healthy cells during time evolution in constant dose-rate ($\dot{D}$) from 0 UAD/step (Unit of Absorbed Dose) to 50 UAD/step. Thin lines correspond to the one standard deviation calculated from the distribution of results obtained at various repetitions of Monte Carlo simulations.

---

[11] Number of constant dose pulses

Another, wider example is presented in Fig. 7, where the fractions of mutated and cancer cells are displayed for many cell colonies, each exposed to different but constant dose-rate. Each point of the curve from Fig. 7 represents a single and independent simulation. In each simulation the results of one cell colony irradiation in the specific dose-rate conditions (horizontal axis) during 300 time steps are presented. Each colony initially consists of $N$ = 1000 healthy cells and results change due to the value of the dose-rate. The results show generally non-linear dose-response in the low dose region and the sigmoidal saturation at high doses (Fig. 7).

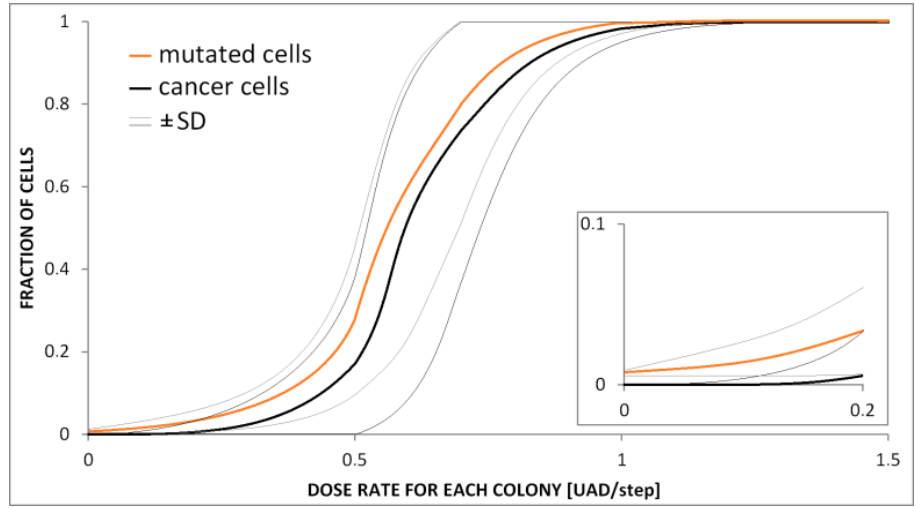

Figure 7. The relationship between the fraction of cells (mutated and cancerous) and dose-rate $\dot{D}$, where each point on the curves represents one colony cultivated in constant value of dose per step during 300 steps starting from $N$ = 1000 healthy cells. Cancer cells are also treated as mutated ones. Thin lines corresponds to one standard deviation (SD). The insert presents the situation below $\dot{D}$ = 0.2 UAD/step. See analogical Fig. 11 in (Fornalski *et al.* 2011a).

The relative ratios of the mutated (orange line) and cancer cells (black line) to all the cells are shown in Fig. 7. As shown, approx. 1% of the cells were initially naturally mutated, i.e. independently of any radiation, after 300 time steps. The linear-quadratic line can be easily drawn through most of the data at low dose-rates, however, the whole relationship is sigmoidal. The potential threshold for cancer cells development appears in low dose region ($\dot{D}$ < 0.1 UAD/step) while for medium dose-rates ($\dot{D}$ = 1 UAD/step) practically all the cells are mutated and neoplastically transformed. Notice also that above $\dot{D}$ = 0.7 UAD/step the total number of cells increases substantially: there are about 5000 cells at $\dot{D}$ = 1 UAD/step and about 27,000 at $\dot{D}$ = 1.5 UAD/step. This effect is caused by fast reproduction of cancer cells ($P_{CS}$). With the still increasing dose-rate, the probability of the cell's death overcomes the probability of its multiplication. Therefore the total number of cells (~100% cancerous) is decreasing when high dose-rate increases ($\dot{D}$ > 20 UAD/step), which is connected with cell killing by high radiation. This effect is essentially used in radiotherapy (discussed later).

### *4.2. Radiation varying over time*

So far the results were based on the constant dose-rate for each cells colony (Fig. 6 and 7). In general, the model allows for a change of the dose in each time step, especially one can implement dose pulses modeling to the group of cells. Such an example is presented in Fig. 8, where situations with small (Fig. 8a), big (Fig. 8b) and mixed (Fig. 8c) dose pulses are presented. In the case of Fig. 8a, not too significant growth of the mutation frequency is observed after the pulse. When the pulse is large (Fig. 8b) one may observe increasing of mutation frequency. What is more important, when same dose is preceded by the smaller one (priming dose) (Fig. 8c) it does not result in the statistically significant change of mutation frequency as compared to $D = 3$ UAD only. This is due to the fact that the adaptive response signal with input data from Table II was too weak.

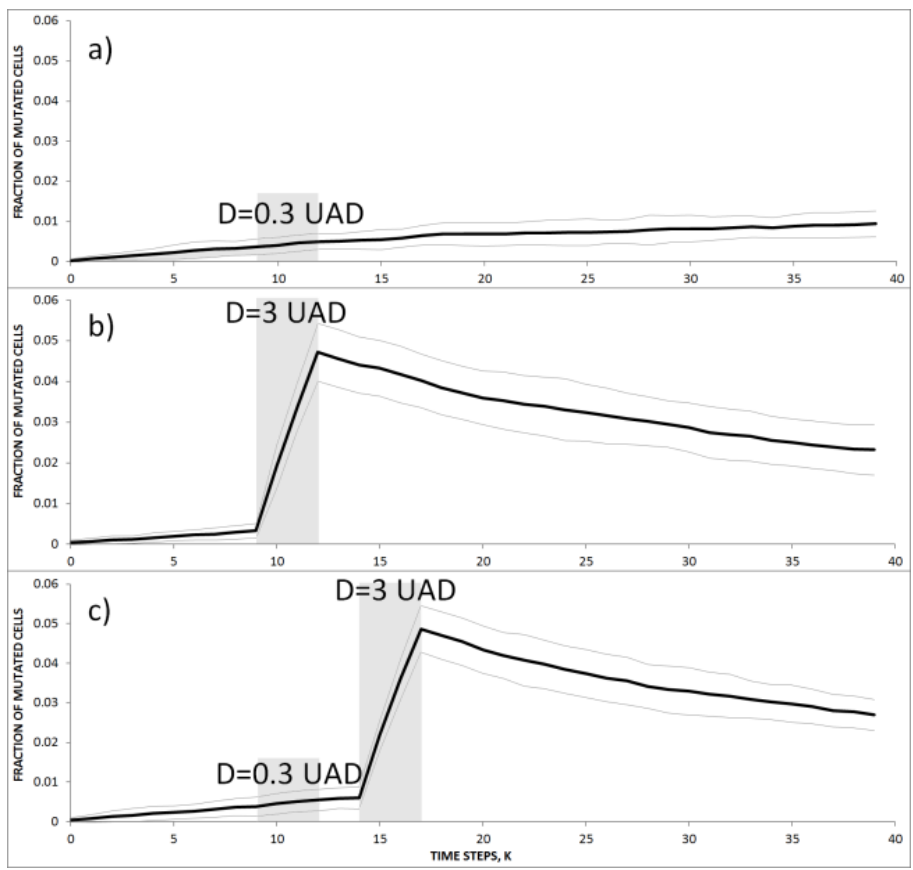

Figure 8. Time evolution of cells in a single cells colony initially consisting of $N = 1000$ healthy cells exposed to dose pulses of a) $D = 0.3$ UAD (Unit of Absorbed Dose), b) $D = 3$ UAD and c) $D = 3.3$ UAD divided into two pulses (with priming and main doses) separated by two time steps. Thin lines display one standard deviation.

### 4.3. Radiotherapy

The idea of using pulses of dose can be easily used in the context of radiotherapy (Enderling and Rejniak 2013). As pointed out in former subsections, according to input conditions from Tab. II, the cancer cells death is possible for dose rates exceeding 20 UAD/step.

The simulated situation with $N$ = 1000 cancer cells at the beginning, is presented in Fig. 9. One can clearly see that such a colony grows rapidly till the radiation pulse 100 UAD/step, applied in 10th time step and lasting through next 10 steps. However, that pulse is insufficient for killing of all cancer cells (Fig. 9a). It seems that only above $\dot{D}$ = 250 UAD/step the cells are killed efficiently (Fig. 9b). For $\dot{D}$ = 250 UAD/step statistically only the 0.25 (0.0 – 0.83, 68% CI[12]) cancer cells survive after the 21th time step. In the last example of $\dot{D}$ = 500 UAD/step, all cancer cells are killed by radiation, which is a goal of any radiotherapy (see also the next section).

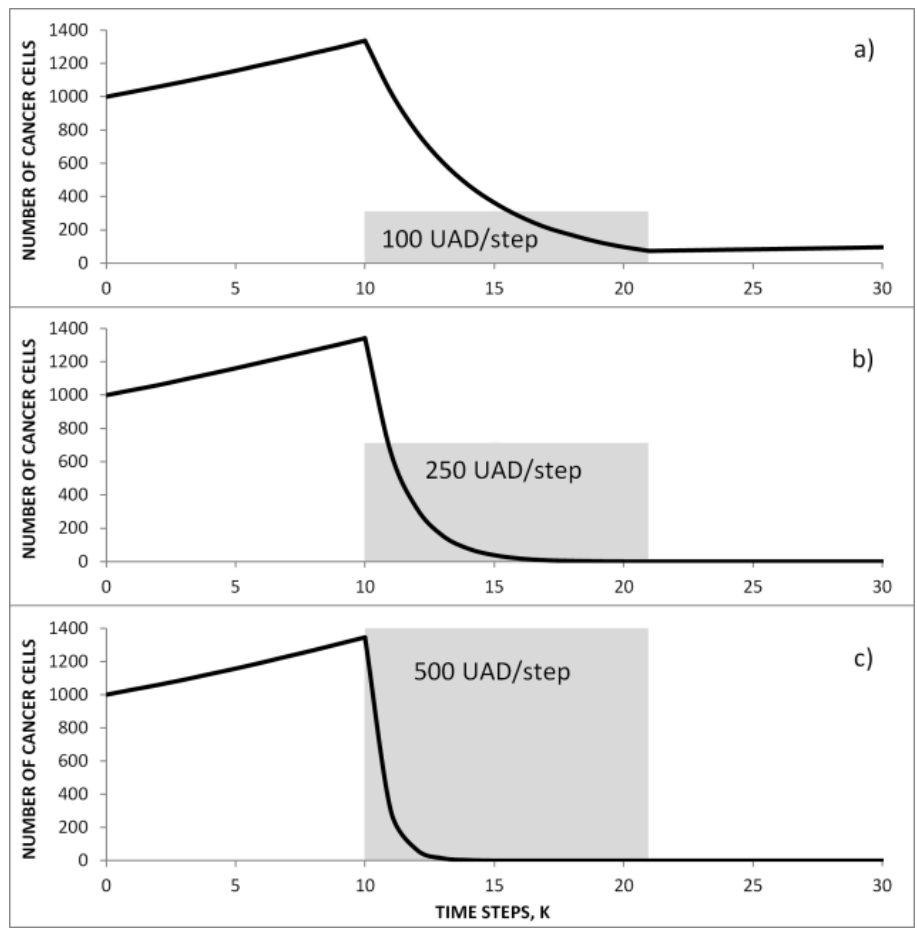

Figure 9. Killing the cancer cells by the radiation pulse applied to a single cells colony initially containing $N$ = 1000 cancer cells. The time evolution of the process is shown for dose rates of a) $\dot{D}$ = 100 UAD/step (Unit of Absorbed Dose), b) $\dot{D}$ = 250 UAD/step and c) $\dot{D}$ = 500 UAD/step.

---

[12] Confidence intervals

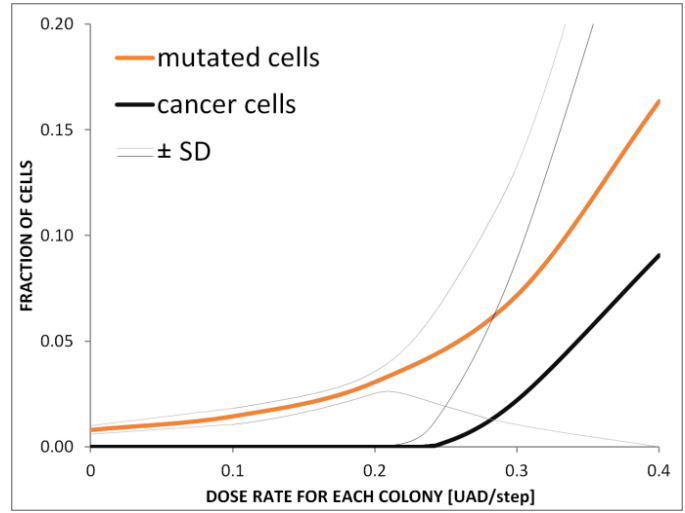

Figure 10. The relationship between the fraction of cells and dose rate $\dot{D}$. It is the same relationship like in Fig. 7 narrowed to 0.4 UAD/step and with the set of parameters $\alpha_1 = 15$, $\alpha_2 = 20$ and $\alpha_3 = 1$ of adaptive response mechanism. Thin lines correspond to one standard deviation (SD) because of many Monte Carlo iterations.

### 4.4. Radioadaptation

The adaptive response mechanism (eq. (21)) with $\alpha$ parameter as chosen in Tab. II is very weak and practically insignificant in complex situations like those presented in Fig. 7 and 8. On the other hand, by changing $\alpha$ parameters it is possible to make the adaptive response mechanism much stronger.

For example parameter values of $\alpha_1 = 15$, $\alpha_2 = 20$ and $\alpha_3 = 1$ move the threshold of cancer cells appearance in Fig. 7 to higher dose rates (Fig. 10). Also the standard deviation margins of mutated cells frequency expand (Fig. 10).

Other interesting views on the time-dependent adaptive response effect observed in the experiment are presented in publications like (Szumiel 2012; Nowosielska *et al.* 2012, 2010; Cheda *et al.* 2006; Boreham *et al.* 2006). In such investigations the group of cells (or organisms) are irradiated by low, medium or mixed low plus medium doses. The model presented in this paper brings very similar results. Fig. 11 presents 4 simulations under 4 different conditions: weak adaptive response input data from Tab. II (Fig. 11a), adaptive response input data from Fig. 10 (Fig. 11b) as well as strengthened (Fig. 11c) and stretched over time (Fig. 11d) adaptive response signal. In all those cases the results show the mutation frequency per cell in background (zero level of radiation), after 0.3 UAD pulse, after 3 UAD pulse and after 0.3 + 3 UAD pulses (separated by two time steps without radiation), like in Fig. 8. In this last case, one can see the effect of a priming dose of 0.3 UAD which substantially reduce the effect of a large dose, see also (Boreham *et al.* 2006; Sykes *et al.* 2006; Seymour and Mothersill 2006). In the case of Fig. 11a, 11b and 11c, the radioadaptation is practically not significant, while strong adaptive response shown in the case of Fig. 11d results in the significant reduction of mutations.

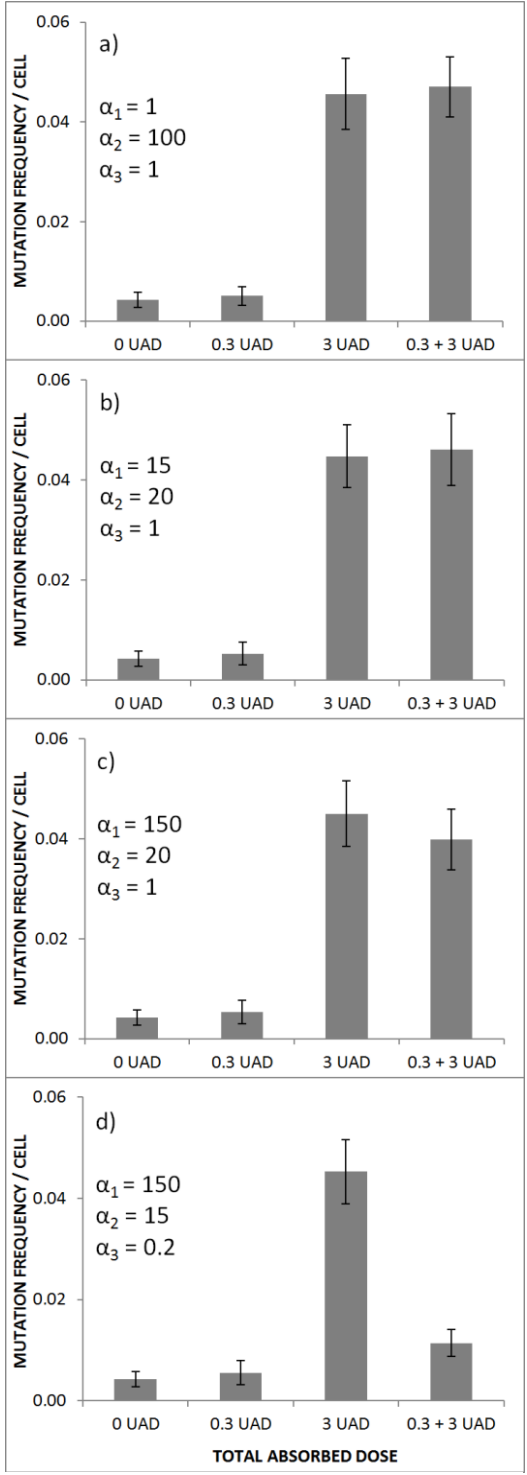

Figure 11. The mutation frequency per cell according to four dose conditions: 0 UAD (spontaneous mutations), 0.3 UAD pulse, 3 UAD pulse and 0.3 + 3 UAD pulses (priming and main doses separated by two radiation-free time steps). The cases from a) to d) correspond to four different adaptive response signals (parameters $\alpha_1$, $\alpha_2$, and $\alpha_3$ in eq. (21))

---

[13] High $K$ causes large probability of mutation in next time step

## 5. DETERMINISTIC RESULTS

### 5.1. Multiplying probabilities

So far, the results were presented when fully stochastic approach was used. However, in some cases there is a need of having some deterministic formulas which can describe in a strict and non-stochastic way the dose-response relationships. Also the presented model can be transformed into the deterministic approach, just by multiplying proper probabilities (Fig. 2), like in (Feinendegen *et al.* 2000, 2010; Leonard *et al.* 2011b). Of course such a procedure can be used for individual needs, for example to obtain the probability distribution of cancer cell's death after single hit by ionizing particle. For example,

$$P_1 = P_{CD} + P_{hit}(P_{RD} + P_{CRD}) \qquad (22)$$

can be used in cells' death simulations connected with radiotherapy. Another example can be connected with a cancer transformation of one cell in one step after single hit by ionizing particle. This will be described by equation:

$$P_2 = P_M P_{RC} + P_{hit} P_{RM} P_{RC} \qquad (23)$$

The visualization of the probability distribution $P_2(D,Q,K)$ is presented in Fig. 12 for three exemplary cases: for the cell with one mutation ($Q=1$), which is either a new one ($K=0$) or old one ($K=100$), as well as for healthy cell ($Q=0$). In this last case it is clearly seen, that the healthy cell, without mutations, cannot be transformed into the cancerous one in single step after single hit by ionizing particle, even when age ($K$) is extremely high[13]. Of course, one cannot exclude that the probability of cancer transformation may increase with age because of the spontaneous mutations ($P_M \cdot P_{RC}$).

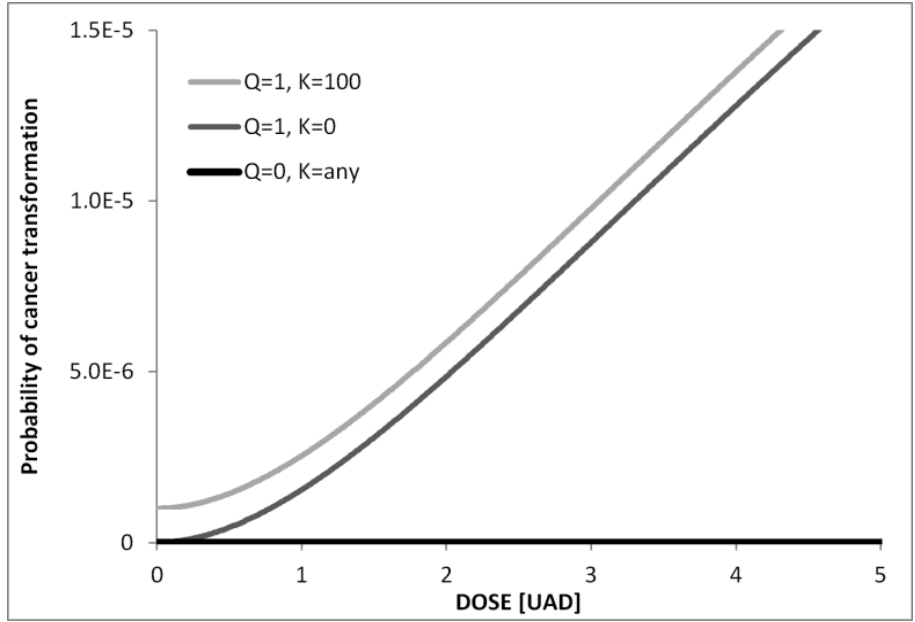

Figure 12. The deterministic probability distribution of cancer transformation in one step after single hit ($D$) by ionizing particle, given by eq. (23), for three exemplary cases.

### 5.2. Balance equations

One of the most important questions given in many deterministic models is how to find the dose point where all positive and negative factors neutralize each other. This point is usually called *NOAEL* (*No Observed Adverse Effect Level*, (Calabrese and Baldwin 1993; Fornalski *et al.* 2012)) and for mutation frequency in the single cell it can be calculated from

$$P_M + P_B + P_{hit}\, P_{RM} = P_R + P_A \qquad (24a)$$

using iterative methods. The left side of eq. (24a) corresponds to all negative effects, where the mutation frequency can increase:

$$Left = (1 - \tau_M) \cdot (1 - e^{-a_M K_{age}^{n_M}}) + \tau_M + \beta_1 (1 - e^{-\beta_2 D_B}) + (1 - e^{-c_{hit} D}) \cdot (1 - e^{-c_{RM} D}) \quad (24b)$$

$K_{age}$ corresponds to the age of cell, $D_B$ means the dose received by neighboring cell(s), while $D$ is searched dose.

The right side of eq. (24a) represents the positive (natural repair and adaptive response) effects which decrease the number of mutations (benefit function (Scott 2013)):

$$Right = \delta\, e^{-a_R K_{age}^{n_R}} + \frac{\alpha_1 \alpha_3^2}{4} D_i^2 K_j^2 e^{2 - \alpha_2 D_i - \alpha_3 K_j} \qquad (24c)$$

It is important to see, that the right side (eq. (24c)) is not dependent directly on the dose actually received, $D$. However, it is dependent on the dose $D_i$ received $K_j$ steps ago. This effect can be called dose inertia.

Fulfilling of eq. (24a) does not mean that the mutation frequency (or the probability that a cell will develop mutation) equals zero. The eq. (24a) represents the *de facto* balance state between creation and depletion of the number of mutations.

The similar reasoning can be used to find the balance equation for cancer cells appearance:

$$P_{CS} + P_{RC}\,(P_M + P_{RM}) = P_{CD} + P_{hit}\,(P_{RD} + P_{CRD}) \tag{25}$$

or, in general form, the balance equation can be written as:

$$PROB\,(negative\ factor) = PROB\,(positive\ factor) \tag{26}$$

Any disturbance in balance equations causes the number of mutations (eq. (24)) or cancer cells (eq. (25)) to increase or decrease. According to all recent knowledge, there is at least one solution of eq. (24) or (25). For example the solution for linear no-threshold (Sanders 2010) model is $D = 0$. For J-shape hormetic model (Sanders 2010) there are two solutions: $D = 0$ and $D = NOAEL$ (eq. (14)). Two solutions exists also for U-shape hormetic model (Sanders 2010): $D = NOAEL_1$ and $D = NOAEL_2$ (where $NOAEL_2 > NOAEL_1$). For the threshold model there is an infinite number of solutions.

### 5.3. Approximations of stochastic results

Curves presented in Figs 6-10 are results of the whole stochastic tree of probabilities (Fig. 2) and only in some cases can be written in a simple deterministic equation, like eq. (22) or (23). But in most of cases a single exact equation does not exist. In such cases it seems to be useful to have a simple approximation of stochastic results, which can be treated as another type of deterministic result.

For example results presented in Fig. 7 can be approximated by a proper sigmoid function (see *Discussion*). Of course such an approximation is only a simplification of a complex tree of probabilities (Fig. 2) but using advanced fittings methods, e.g. Bayesian (Fornalski *et al.* 2010), it is easy to find the best fit to the simulated data.

## 6. DISCUSSION

All results presented in this paper show different processes. One can find simulations of the effect of a priming dose, the behavior of virtual cells in constant dose-rate environment as well as radiotherapy modeling. The deterministic results can omit the entire tree of probabilities (Fig. 2) and focus on the detailed problem only. All those results are in general agreement with the experimental data, but details depend on the input parameters values from Tab. II.

### 6.1. Experimental data

Each model, including the presented one, has to be based on real experimental data. These are usually difficult to obtain because, in reality, there are many types of cells with different properties, age, etc. More than that, the concept of "virtual cell" makes all input parameters (Tab. II) depending on the decision of the scientists who carries out calculations. Sometimes, some of probabilities can easily be found, like *const = 6·10⁻⁵* from $P_{RD}$ is taken from (Feinendegen *et al.* 2010), but typically input parameters to the model have to be measured step by step and cell-type by cell-type.

The most basic relationships, like $P_{RM}$, $P_{RC}$ or $P_M$, can be fitted using the popular relationships between absorbed dose and chromosomal aberrations frequency in blood, *Y(D)*. Sometimes the *Y(D)* relationships are named persistent residual breaks relationships (Scott *et al.* 2007; Scott 2011) or calibration curves (Szłuińska *et al.* 2005; Brame and Groer 2003; Kellerer and Rossi 1974). However, the most popular forms of *Y(D)* are presented for neutron radiation as (Szłuińska *et al.* 2005):

$$Y_n(D) = Y_0 + c_1 D \; \propto \; P_{M+RM}^{(n)} \tag{27}$$

and for gamma radiation (Szłuińska *et al.* 2005):

$$Y_\gamma(D) = Y_0 + c_1 D \; + \; c_2 D^2 \; \propto \; P_{M+RM}^{(\gamma)} \tag{28}$$

where $P_{M+RM}$ is a join probability for $P_M$ and $P_{RM}$ (Tab. II). Both relationships *Y(D)* can easily be determined experimentally, however eq. (27) and (28) cannot be universal ones (because $\lim_{D\to\infty} Y(D) = \infty$). For this reason one can use scaled eq. (1) instead of (27) (Fornalski 2014):

$$Y_n^*(D) = (Y_{max} - Y_0) \cdot (\, 1 - e^{-c_1 D}) + Y_0 \tag{29}$$

and scaled eq. (3) or (5) instead of (28) (Fornalski 2014):

$$Y_\gamma^*(D) = (Y_{max} - Y_0) \cdot (1 - e^{-c_1 D^{c_2}}) + Y_0 \tag{30}$$

$$Y_\gamma^{**}(D) = (Y_{max} - Y_0) \cdot (1 - e^{-c_1 D - c_2 D^2}) + Y_0 \tag{31}$$

The $Y_{max}$ represents the maximum possible number of damages (mutations) *Q* per cell (one can assume $Y_{max}=1$), and $Y_0$ is the natural (non-radiation induced) level of spontaneous damages per cell. Thus $Y_0$ can correspond to $P_M$ and from the experiment its average value is $Y_0 \approx 0.0005$ (Szłuińska *et al.* 2005). Equations (29)-(31) can be alternatively useful in the context of biological dosimetry (Fornalski 2014).

All relationships, presented as input assumptions in Tab. II as well as in whole third section, can be fitted to exact data points. Assuming that such data points are available, one can use many parameter estimation techniques. However, one can expect that such data would be scattered and subjected to systematic errors. Therefore the best method of finding all necessary curves from experimental data is the Bayesian fitting analysis (Fornalski *et al.* 2010; Fornalski and Dobrzyński 2011; Fornalski 2014). This method is recommended whenever experimental data is generally problematic.

### 6.2. Potential modifications of the model

The third section (*Physical Basis and Input Data*) contains several input equations with their physical background. Such equations have their own advantages but can create also some limitations of the model. However, the presented model is flexible and all potential users can use their own curves of probability distributions as well as modify the tree of probabilities (Fig. 2).

There are some propositions of model modifications below, which can be used by potential users.

#### 6.2.1. Repair of damages

In some cases the potential changes are dependent on the cell type, tissue type or methodology. As an example let us look at the possibility of potential change in the probability distribution of damage repair in the cell, $P_R$, given originally by sigmoid equation (4). This assumption is only time dependent ($K$), so the potential change can be connected with the relationship dependent on number of damages (mutations), $Q$. This approach can be found in (Scott 2011) where "*repair of radiation induced DSB (Double Strand Break) is assumed to represent a Poisson process*". The *repair halftime* is proposed as 0.69 $\beta$, where the average repair rate per break is given by $\mu = 1/\beta$ (experimentally found that $\beta$ = 2.5 h). In a similar way, "*DNA DSBs have a dose- and time-dependent Poisson distribution after low to moderate doses of ionizing radiation*" (Scott 2011). Finally, the probability distribution of repair $Q$ damages is given by Erlang distribution (Scott 2011):

$$P_R(K,Q) = \frac{\beta^{-Q} \, K^{Q-1}}{(Q-1)!} \, e^{-K/\beta} \qquad (32)$$

The Erlang distribution (32) has similar property to the Poisson one. More information and results are presented in the original paper (Scott 2011).

#### 6.2.2. Adaptive response

The presented probability density function $P_A$ (eq. (21)) is based originally on equations (16) and (17). They can also be presented in more rapidly varying forms as

$$P(\xi) = c_1 \left(1 - e^{-c_2 \, \xi}\right) e^{-c_3 \, \xi} \qquad (33)$$

which can be potentially used alternatively to eq. (16)-(17). However, the form (33) gives results of adaptive response influence similar to (16)-(17).

The adaptive response phenomena is usually not included into the radiation protection standards (Feinendegen *et al.* 2010), where LNT hypothesis (*Linear No-Threshold*) is used for cancer risk estimation (which is assumed to be proportional to accumulated dose, $D$). The LNT model can be presented in popular form of

$$P_{risk} = c_1 D \qquad (34)$$

and taking eq. (34) as an assumption, one can instinctively include simplified (time independent) adaptive response to LNT model as:

$$P_{risk} = c_1 D \ (1 - e^{-c_2 D}) \tag{35}$$

However, such models presented by eq. (34) and (35) have a mathematical disadvantage, because $\lim_{D \to \infty} P_{risk} = \infty$. To avoid this problem, one can use more natural and saturated eq. (1) instead of eq. (34) and thanks to that eq. (35) would be a sigmoid one (eq. (5)-(6)) or even hormetic (eq. (13)).

### 6.2.3. Bystander effect and BaD model

The form of probability distribution describing bystander effect was proposed by (Prise *et al.* 2003) and is implemented in present model as eq. (9). However, in some literature (Brenner *et al.* 2001) one can find different dependences, similar to equations (16) or (33).

The form of (33) was used in *BaD model* proposed by (Brenner *et al.* 2001) and described also in (Leonard 2008; Leonard *et al.* 2011a). In this approach one can find the joint probability distribution for bystander effect, $P_B$, and radiation induced mutations, $P_{RM}$, as ($\xi = D$):

$$P_{RM+B}^{(BaD)}(\xi) = c_1 \xi + c_2 (1 - e^{-c_3 \xi}) e^{-c_4 \xi} \tag{36}$$

The equation (36) can safely be used at low values of doses ($\xi = D$) only. When the dose increases, the mathematical disadvantage appears, because $\lim_{\xi \to \infty} P(\xi) = \infty$. It is because the first term of eq. (36) is the linear form of (34), which can be changed to quasi-linear saturated form of eq. (1). Therefore it is reasonable to use:

$$P_{RM+B}(\xi) = P_{RM} + P_B^{(BaD)} = 1 - e^{-c_1 \xi} + c_2 (1 - e^{-c_3 \xi}) e^{-c_4 \xi} \tag{37}$$

which does not differ from (36) at low doses but saturates at high doses. This modification of *BaD* assumption can also be potentially used in presented model. One of its property, the "hunchbacked" shape, can also be useful in the discussion about the potential radiation hypersensitivity (Prise *et al.* 2005).

### 6.2.4. Distance dependent bystander effect

The model presented in this paper assumes that the bystander effect probability distribution is dose-dependent only (eq. (9)). This simplification can be accepted in current situation because the whole virtual organism of cells is computationally arranged as a one-dimensional multivariable chain[14] (Fig. 1). But in a more general situation, the cells can be spatially arranged in two- or three dimensional matrix, which requires careful accounting on the distance ($r$) from the irradiated cell. The situation, where the probability of bystander effect appearance decrease with the distance, can be well described by the Poisson distribution connected with single hit (Gaillard *et al.* 2009; Sasaki *et al.* 2012):

$$P_B(\text{r}) = c_1 \frac{e^{-\lambda} \lambda^r}{r!} \equiv \frac{1}{r!} \tag{38}$$

simplified for expected value equals *λ=1* and normalized to *P(r=0)=1*.

---

[14] However, in the recent version of the model only bystander effect creates interactions between cells. So the term "cells' colony" should be understood just as a collection of quasi-independent cells only; one-dimensional is the chain, not the colony

Using similar reasoning like in the case of eq. (19), one can postulate the distance and dose (*D*) dependent probability distribution of bystander effect as:

$$P_B(\text{D,r}) = \frac{\beta_1}{r!} \left(1 - e^{-\beta_2 D}\right) \tag{39}$$

for dose-dependent eq. (9), or:

$$P_B^*(\text{D,r}) = \frac{\beta_1}{r!} \left(1 - e^{-\beta_2 D}\right) e^{-\beta_3 D} \tag{40}$$

for dose-dependent equation from *BaD* model (eq. (33)). Those propositions given by eq. (39) and (40) can be considered for the potential development of the model.

### 6.2.5. *Matrix arrangement of virtual cells*

The next potential change in the model development was introduced in previous point, namely all quasi-independent cells are arranged now as a one dimensional multivariable chain (Fig. 1). However, in a more general situation the cells can be arranged in two- or three dimensional matrix. Thanks to introducing the 3D matrix of cells, it would be easy to irradiate the selected cells only (radiation microbeam) without irradiation of the whole colony. This may enable the simulation of such effects as cell-to-cell interactions, the effect of cell cycle, secondary cancers, radiotherapy of tumor surrounded by non-transformed cells or even the potential use of Bragg curve in heavy ion irradiation. Such a modification can also be very useful in the case of microdosimetry introduction into the model.

### 6.2.6. *Many mutations after single hit*

One more potential change can also be connected with the elimination of one inconvenience presented indirectly in Fig. 12. The recent assumption is that during a single time step a single cell can receive a maximum of one mutation only. In general, there can be many mutations after single high dose pulse irradiation. It is expected that the probability distribution connected with mutations appearance (*Q*) after single hit in single cell is given by Poisson distribution, which can significantly improve the model during future development.

Many potential future modifications of the model were presented above. The model is a universal tool and its further development is possible.

### 6.3. Generality of all input equations

Quick look at all relationships presented in Tab. II suggests that all of them can be written in one wider form, like

$$P(\xi) = P_{min} + (P_{max} - P_{min})\left(a - b\xi^n e^{\sum_{k=1}^K c_k \xi^{m_k}}\right) \tag{41}$$

or, in the most general form:

$$P(\xi) = P_{min} + (P_{max} - P_{min})\sum_{i=1}^I \left[\prod_{j=1}^J \left(a_{i,j} - b_{i,j}\xi^{n_{i,j}} e^{\sum_{k=1}^K c_{i,j,k}\xi^{m_{i,j,k}}}\right)\right] \tag{42}$$

Equation (42) is a universal one for all relationships presented in the model. Taking all variables ($a_{i,j}$, $b_{i,j}$, $n_{i,j}$, $c_{i,j,k}$ and $m_{i,j,k}$; $P_{min}$ and $P_{max}$ are just a scaling factors) as proper values, one can acquire all presented response curves, like Avrami's, hormetic, quasi-linear or *BaD* model's curve. For example for $I=J=K=1$, $b_{1,1}>0$, $n_{1,1}=0$, $a_{1,1}=1$, $c_{1,1,1}<0$ and $m_{1,1,1}>1$ the eq. (42) becomes the Avrami equation (3). Changing $m_{1,1,1}$ to be equal 1, the eq. (42) becomes a quasi-linear function of eq. (1). Similarly, for $K=2$, $c_{1,1,1}<0$, $c_{1,1,2}>0$, $m_{1,1,1}>1$ and $m_{1,1,2}=1$ one acquire the hormetic curve of eq. (13).

The possibility of presenting all dose-response relationships by one general equation (42) is a real simplification and can be potentially used to compare results with experimental data.

### 6.4. Natural nonlinearity of complex systems

Laws of physics take priority over biological laws. For much the same reason, the mathematical laws take priority over physical ones. This is also the reason why mathematical reasoning should be implemented into biophysical one.

For example, let us look at the single particle as a simple non-complex system. When such a particle is hit by a second particle (ionizing radiation), the response is linear according to the conservation of momentum, energy and mass. When there is not a single particle but many particles (like in DNA chain), the problem is more complex. For example the neutron interaction (Fig. 13a) due to the simple kinetics creates the linear response (eq. (27)). In the case of photon interaction the problem is more complicated due to Compton scattering and photoelectric effect (Fig. 13b). The additional dependence based on the value of radiation LET (linear energy transfer) can make the linear-quadratic response of chromosomal aberration frequency (eq. (28)), which was postulated in (Kellerer and Rossi 1974; Schmid *et al.* 1998). More complex systems, like cells (Fig. 13c), cannot have linear response, in spite of linear relationships given as inputs.

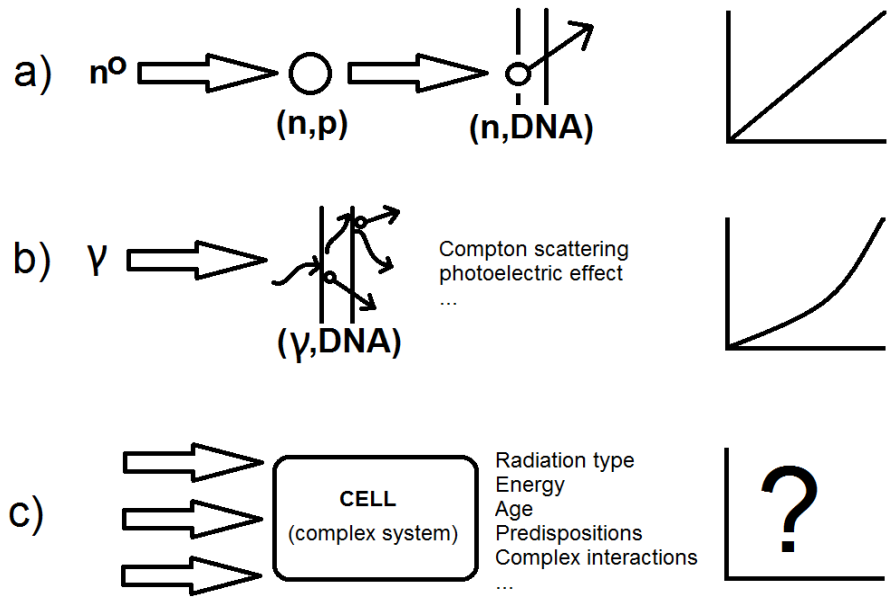

Figure 13. The simple examples of potential interactions with radiation (indirect effects excluded): a) DNA interaction with neutron radiation, b) DNA interaction with gamma radiation and c) cell (as a physical complex system) interaction with mixed radiation.

In general, the organism, even a single cell, can be treated as a complex system. In that system the multiplication of many input parameters, plethora of internal conditions and dependences, makes the natural non-linear dose-response of the organism. This phenomenon is just a simple cause of mathematical laws in a biophysical world.

The results of presented model confirm that where multiple relationships linearly depending on the dose are introduced, the overall reaction generally shows a non-linear response.

## 7. CONCLUSIONS

The stochastic model used in the present paper shows what potentially can happen in a population of hypothetical generalized cells exposed to ionizing radiation. Such a stochastic approach seems to be better than many deterministic models because it takes into consideration some individual susceptibilities and probabilities rather than many phenomenological factors which, although called "*probabilities*" are not, in fact, used as such. In the present model, every run yields somewhat different results only because of the use of the truly probabilistic approach[15]. However, the obtained results are consistent with many epidemiological and experimental data demonstrating or implicating a threshold and/or sigmoidal shape of the response. It is crucial to the success of the presented approach that primary damage to an essential target, such as the DNA, is analytically separated from the response of the whole system to that damage.

---

[15] Which is close to real situations, where everything is described by some probability distributions and random processes; thus, the final result can be presented as an average value of many different single simulations, which arrange into some Gaussian-like distribution

One of the advantages of using of a probability tree is a possibility of easy addition or cut off some of its branches. The present algorithm has been significantly simplified in terms of its biological background, but it can certainly be further developed and made more complex in the future. Also the general and not precisely defined unit of absorbed dose and dose-rate is used to keep a wide point of view. What is of interest is whether a model developed based on the hypothetical responses of virtual cells to hypothetical irradiation can be successfully applied to real world data such as has been obtained for effects like radiation-induced mutation, neoplastic transformation, tumor cell killing, etc.

Most of input relationships are quasi-linear ones (eq. (1)), but some of them, especially the probability of cancer transformation, $P_{RG}$, are sigmoidal according to theories of nucleation and growth or catastrophe. The sigmoidal shape of organism response is often found in the literature (Tubiana *et al.* 2011; Zyuzikov *et al.* 2011; Henriksen and Maillie 2003; Dabrowski and Thompson 1998; Haskin *et al.* 1997; Anderson and Storm 1992; Moolgavkar and Luebeck 1990; Maisin *et al.* 1983; Ashley 1969; Armitage and Doll 1957) and seems to be natural for deterministic effects of radiation. However, the presented model shows that the similar relationship can also describe stochastic effects (Fig. 7). It is the potential significant modification of the linear (LNT) assumption for stochastic (cancer) effects used worldwide so far.

In the present model an undefined cell culture with no association to any specific tissue has been used. Indeed, the present model has no ambition to develop the algorithm for more complex tissue reactions. However, a potential user of the model can take their own values of input parameters specifying the type of cells or the tissue they may want to investigate, so the presented model should be treated just as a tool. It can be applied to simulate cancer transformation, adaptive response effect, radiotherapy and many others. More than that, the presented paper contains some new concepts like dose- and time-dependent adaptive response equation (21), balance equation (26) or general dose-response equation (42). However, the model has also many limitations due to the very simplified biology of such virtual cells.

The present approach may provide clues for modeling the impact of low doses on organisms with use of the quasi-Markov chain Monte Carlo techniques with physical background. A prospective user of the model can add new branches to the probability tree and/or can select a completely different set of the input probabilities. The values of such parameters listed in Tab. II are just an example of how the model can work. In order to describe a real biological system, the model will be further developed and elaborated in the future and will be applied to real situations.

## ACKNOWLEDGEMENTS

Author wish to thank Prof. Ludwik Dobrzyński from National Centre for Nuclear Research (Otwock-Świerk, Poland) for the motivation, inspiration, carefully reading the manuscript and providing many valuable remarks and discussions. Additional thanks to Mr. Michał Kurpiński (PGE EJ 1) for linguistic corrections.